\documentclass[a4paper,11pt]{article}
\usepackage{pos}

\renewcommand{\hookAfterAbstract}{%
  \par\bigskip
  \textsc{P3H-24-060, TTP24-032}
  \href{}
}

\title{Towards NNLO QCD corrections to Higgs boson pair production}

\author*[a]{Matthias Steinhauser}

\affiliation[a]{Institut f{\"u}r Theoretische Teilchenphysik,\\
   Wolfgang-Gaede Stra\ss{}e 1, 76128 Karlsruhe, Germany}

\emailAdd{matthias.steinhauser@kit.edu}

\abstract{
Higgs boson pair production plays an important role in the determination of
the Higgs boson self coupling. The
predictions based on next-to-leading order corrections show a large dependence
on the renormalization scheme of the top quark mass, which requires a
next-to-next-to-leading order calculation. We discuss the current status and
show first results of the
three-loop virtual corrections.
}

\FullConference{42nd International Conference on High Energy Physics (ICHEP2024)\\
18-24 July 2024\\
Prague, Czech Republic\\}

\begin{document}
\maketitle


\section{Introduction}

Higgs boson pair production is an active field of research, both on
the experimental side but also in the theory community.  It is
probably the most promising observable to learn more about the Higgs
boson self coupling, a crucial element in the Standard Model Lagrange
density.

Theory predictions for Higgs boson pair production are quite advanced; due to
size limitations not all references can be listed in these proceedings and
we concentrate on those which are most relevant for the results presented
below.  The leading order (LO) corrections were computed long before
the discovery of the Higgs boson~\cite{Glover:1987nx,Plehn:1996wb}.  About
25~years ago the first next-to-leading order (NLO) result became
available, however, only in the limit of infinitely heavy top
quark~\cite{Dawson:1998py}.  Exact NLO results have been known since
2016~\cite{Borowka:2016ehy,Borowka:2016ypz,Baglio:2018lrj}. To obtain these
results, to a large extent numerical methods have been used, which have several
drawbacks.  For example, there is limited flexibility in connection to
renormalization scheme changes and, furthermore, the calculations require a
fair amount of CPU time for the numerical evaluation of the cross section.

In Section~\ref{sec::nlo} we propose an alternative to the (mostly) numerical
approach: analytic expansions. Here the idea is to identify, in a given region
of phase space, small parameters and then perform a Taylor or an asymptotic
expansion.  For Higgs boson pair production first steps in this direction have
been undertaken in Ref.~\cite{Davies:2019dfy} where analytic results
from~\cite{Davies:2018qvx} have been used in the high-energy region and
(expensive) numerical results from~\cite{Borowka:2016ehy,Borowka:2016ypz} are
only needed in a restricted phase space. A further development of this idea
was realized in Ref.~\cite{Bellafronte:2022jmo} where the expansion around the
forward-scattering limit~\cite{Bonciani:2018omm} has been combined with the
high-energy results~\cite{Davies:2018qvx} abandoning completely the need for
numerical evaluations.  A different formulation of this idea can be found in
Ref.~\cite{Davies:2023vmj}, where many more expansion terms have been used in
the different limits. This leads to precise results in the whole phase space.
The results of Ref.~\cite{Davies:2023vmj} are summarized in
Section~\ref{sec::nlo}.

In Ref.~\cite{Baglio:2020wgt,Bagnaschi:2023rbx} it has been shown that the NLO
prediction for Higgs boson pair production suffers from large
uncertainties originating from the renormalization scheme for the top quark
mass. In order to tame these uncertainties it is necessary to perform a
calculation at next-to-next-to-leading order (NNLO). Until recently NNLO
effects were basically only available in the large-$m_t$
limit~\cite{deFlorian:2013jea,deFlorian:2013uza,Grigo:2014jma}. The most
advanced calculation has been performed in Ref.~\cite{Davies:2019xzc} where five
expansion terms in $1/m_t$ for the box form factors for $gg\to HH$ have been
computed. In the meantime first results for other kinematic regions are
available: In Ref.~\cite{Davies:2023obx} the light-fermion contribution to the
form factors has been computed for vanishing transverse momentum of the Higgs
bosons and in Ref.~\cite{Davies:2024znp} results valid in the whole phase
space for all reducible contributions have been obtained.  The results of
Refs.~\cite{Davies:2023obx} and~\cite{Davies:2024znp} are summarized in
Section~\ref{sec::nnlo}.


\section{\label{sec::nlo}NLO}

In Ref.~\cite{Davies:2023vmj} it was shown that for the NLO QCD
corrections to $gg\to HH$ it is possible to cover the whole phase
space by combining expansions in the forward scattering kinematics and in the
high-energy region.

We have implemented the forward scattering expansion by performing a Taylor
expansion of the integrand in the Mandelstam variable $t$ up to order $t^5$
and in the external Higgs boson mass $m_H$ up to quartic order.  A similar
approach has be used before in Ref.~\cite{Bellafronte:2022jmo}. However, fewer
expansion terms were incorporated in the approximation leading to less
precise results for the two-loop virtual corrections.

In the high-energy region we have constructed an expansion up to $m_t^{112}$
and up to $m_H^4$.
To obtain precise results it is crucial to accompany the deep
expansion with a Pad\'e approximation. In our case this
leads to precise results, even for quite small values
of $p_T$. Furthermore, our approach also provides an
estimate of the uncertainty. For details concerning the
Pad\'e procedure we refer to Ref.~\cite{Davies:2020lpf}.

\begin{figure}[t]
  \begin{center}
  \begin{tabular}{cc}
    \raisebox{1em}{\includegraphics[width=0.4\textwidth]{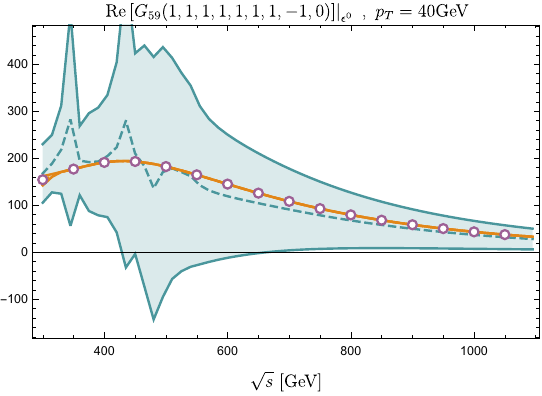}}
    &
    \includegraphics[width=0.45\textwidth]{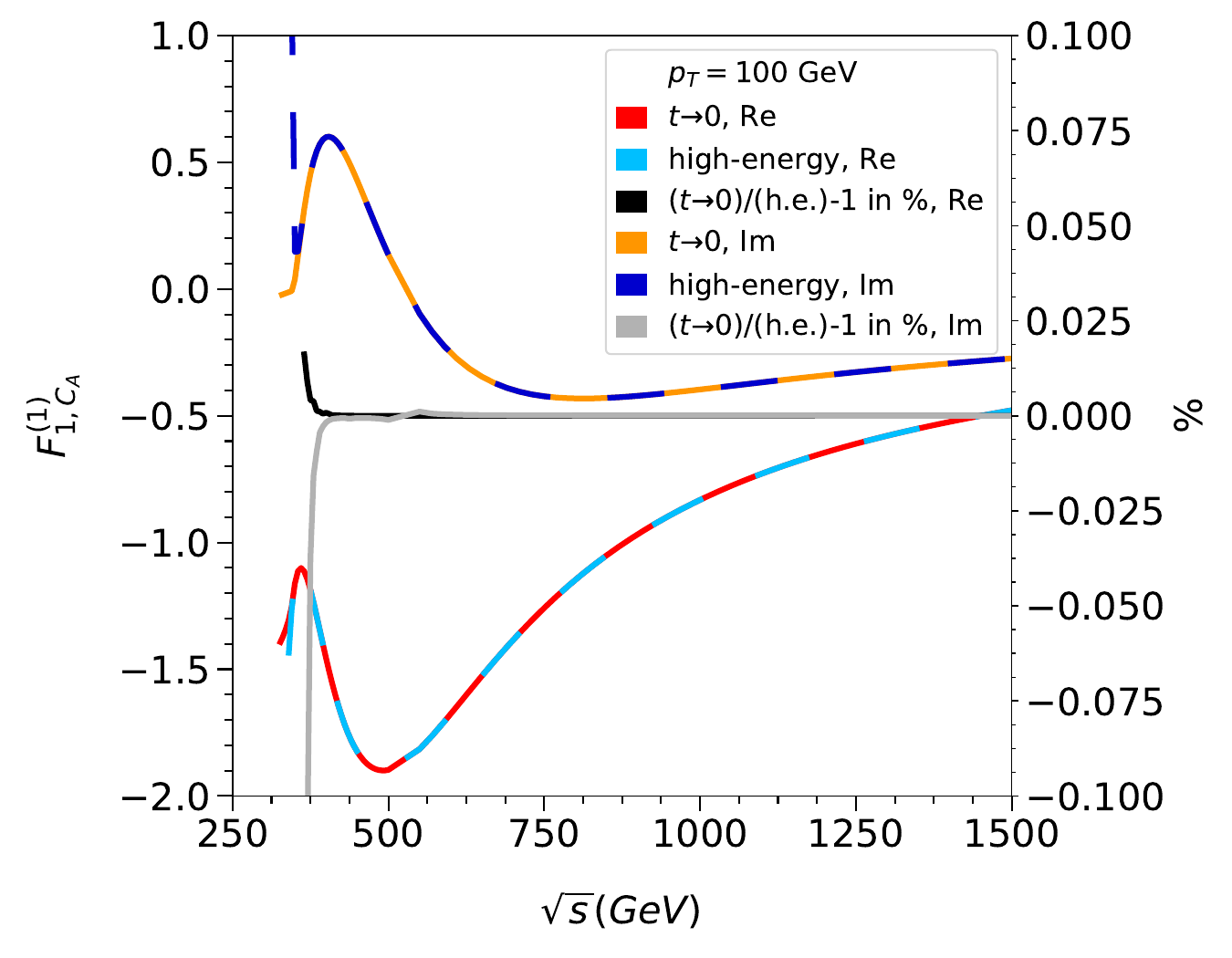}
    \\
    (a) & (b)
  \end{tabular}
  \end{center}
  \caption{(a) Comparison of Pad\'{e}-based approximations constructed from
    different expansion depths (see text) with numerical
    results obtained using {\tt FIESTA}, for a non-planar master integral with
    a numerator.  (b) $C_A$ contribution to the two-loop form factor
    $F_{\rm box1}^{(1)}$ as a function of $\sqrt{s}$ for $p_T=100$~GeV.}
  \label{fig::non-pl_MI}
\end{figure}

In Fig.~\ref{fig::non-pl_MI} (taken from Ref.~\cite{Davies:2023vmj}) we
show results of a non-planar scalar integral (see Fig.~1 of
Ref.~\cite{Davies:2023vmj}) as a function of $\sqrt{s}$ for fixed transverse
momentum $p_T=40$~GeV. Let us stress that from the perspective of the
high-energy expansion this is a quite small value.  Exact results obtained
with {\tt FIESTA}~\cite{Smirnov:2021rhf} are shown as open circles.  The
greenish line and band correspond to results where an expansion up to
$m_t^{32}$ has been used as input. It is interesting to note that the central
line is quite close to exact numerical results. Furthermore, the uncertainty
estimate is reliable.  The orange line and band are obtained using an
expansion up to $m_t^{112}$. The band is only visible close to the threshold
for top quark pair production.  The orange curve reproduces the exact result
with impressive accuracy which demonstrates that in our application the Pad\'e
approximation is a precision tool.

For the $t$ expansion no Pad\'e improvement is necessary since we observe a
fast convergence in the region of phase space where we apply this expansion
(see below).

In Fig.~\ref{fig::non-pl_MI} we show results for one of the colour structures
of the (renormalized and infrared-subtracted) two-loop form factor $F_1$ (see
Ref.~\cite{Davies:2023vmj} for a precise definition) for fixed $p_T=100$~GeV
as a function of the partonic center-of-mass energy, $\sqrt{s}$. The red and
light blue curves show the real part and the orange and dark blue curves the
imaginary part.  In both cases we plot the results from the high-energy and
the forward scattering approximation.  The gray curves show relative
differences in percent (see right axis).  For the chosen value of $p_T$ one
observes an agreement far below the per mille level. Similar results are
obtained for a wide range of $p_T$ between
$50~\mbox{GeV}\lesssim p_T \lesssim 200~\mbox{GeV}$ which defines a
comfortable overlap region where both approximations are valid.  For larger
values of $p_T$, i.e. higher energies, the high-energy expansion is supposed
to work even better.  On the other hand, for smaller $p_T$ we enter deeper
into the regime of the forward scattering approximation.  Thus, we can cover
the whole phase-space based on analytic expansions and without relying on
expensive numerical evaluations.  The limiting factor of this approach on the
final precision is the expansion in $m_H$. Including quartic terms, the
corresponding uncertainty can be at the percent level if $p_T$ and $\sqrt{s}$
are small.


\section{\label{sec::nnlo}NNLO}

\begin{figure}[t]
  \begin{tabular}{cccc}
  \includegraphics[width=0.24\textwidth]{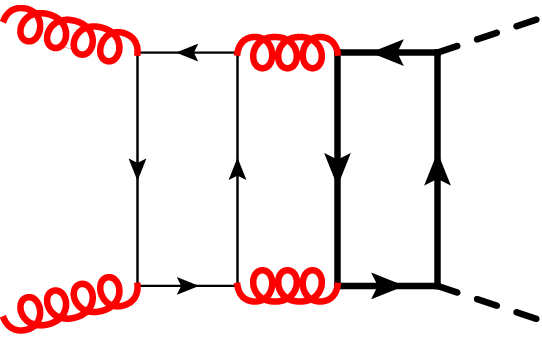}  &
  \includegraphics[width=0.16\textwidth]{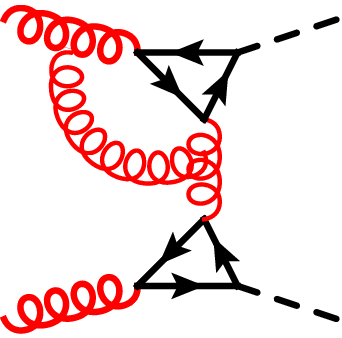} & 
  \includegraphics[width=0.24\textwidth]{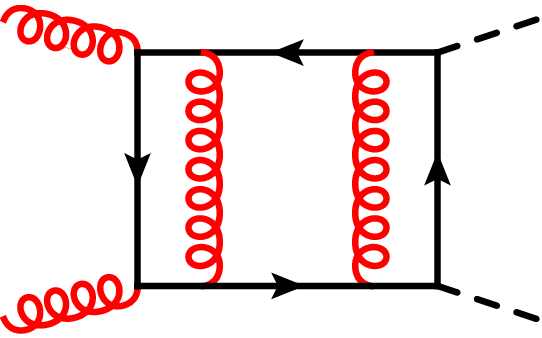}  &
  \includegraphics[width=0.24\textwidth]{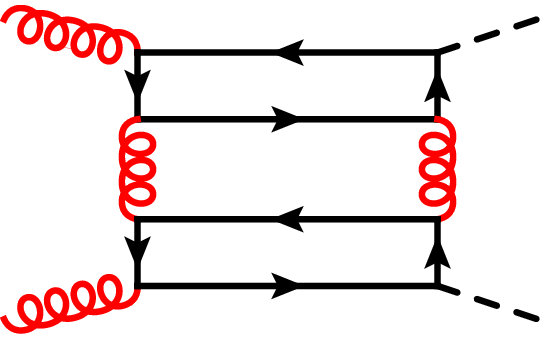} 
    \\ (a) & (b) & (c) & (d)
  \end{tabular}
  \caption{\label{fig::3loop_class} Classification of the three-loop
  virtual corrections to $gg\to HH$.}
\end{figure}

The NNLO contribution which is sensitive to the top quark mass scheme used in 
the NLO prediction are the three-loop virtual corrections.  This requires the
calculation of triangle and box diagrams with internal and external massive
particles.  The triangle contributions have been computed in
Refs.~\cite{Davies:2019nhm,Davies:2019roy,Harlander:2019ioe,Czakon:2020vql,Niggetiedt:2023uyk} which leaves us with the box diagrams.  In
Fig.~\ref{fig::3loop_class} we show a classification of the Feynman diagrams
which will be discussed in the following in more detail.

Among the methods used at two loops, in our opinion, the most promising
approach for the three-loop calculation is the forward-scattering expansion.
Here, one can exploit the fact that the expansion in $t$ is a Taylor
expansion.\footnote{This is not true for the diagram in
  Fig.~\ref{fig::3loop_class}(d), see below.} Thus one can expand the
integrand before the reduction to master integrals yielding dependence only on
the scales $s$ and $m_t^2$. The other methods rely on IBP reductions involving
more scales which are currently not feasible.

\begin{figure}[t]
  \begin{tabular}{cc}
  \includegraphics[width=0.45\textwidth]{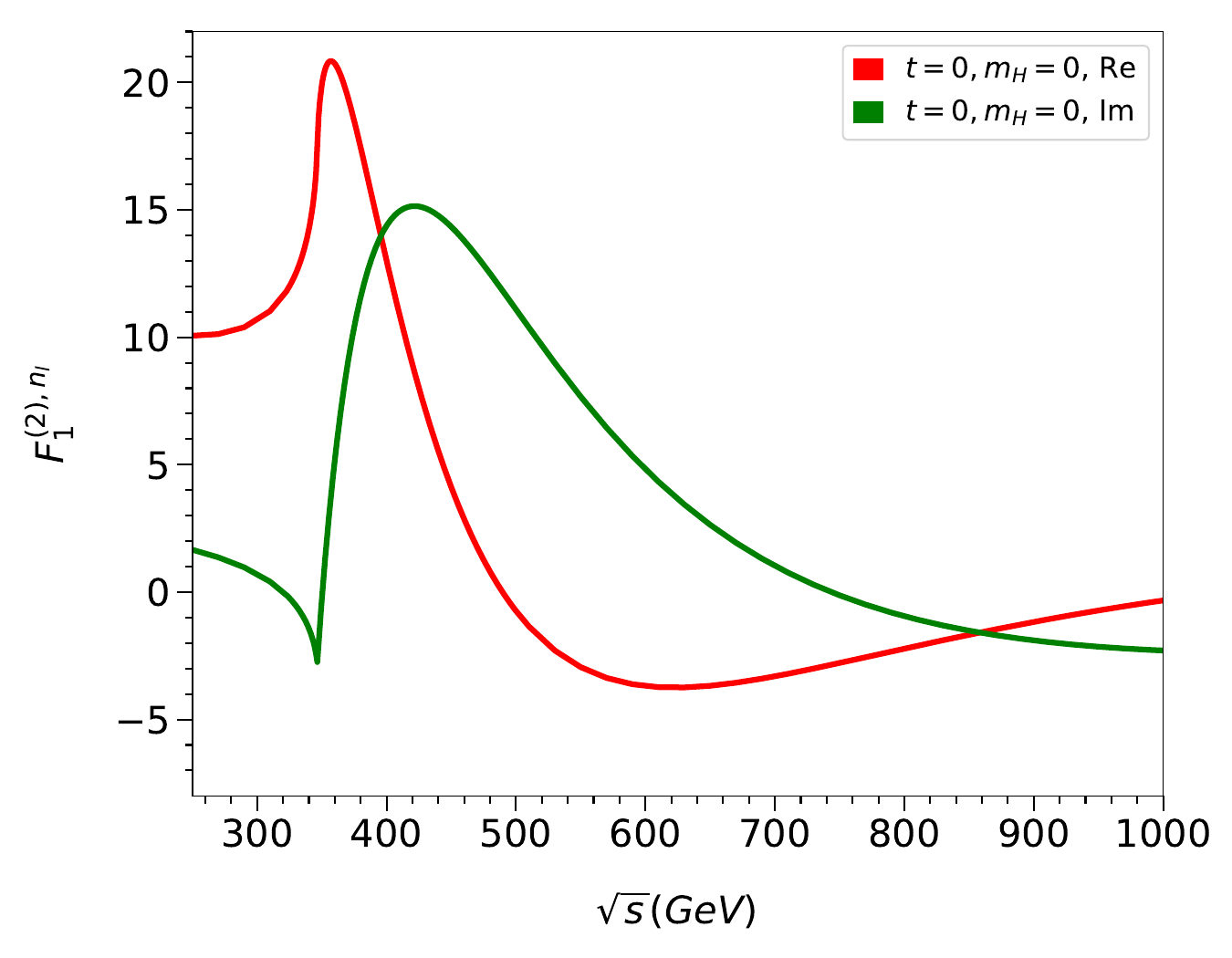}
  &
  \includegraphics[width=0.45\linewidth]{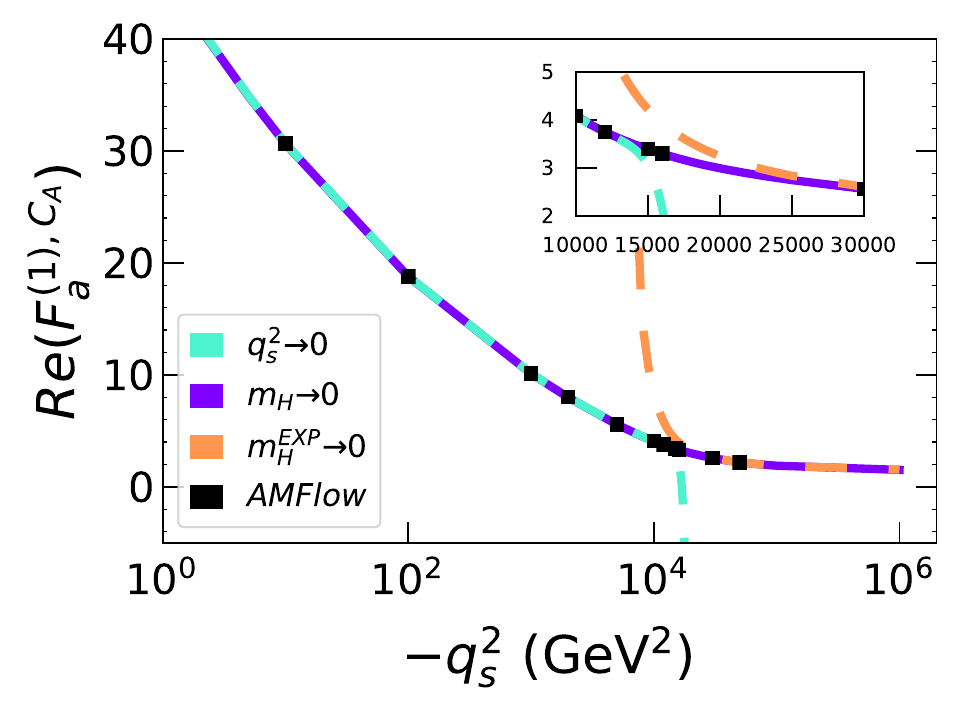}
  \end{tabular}
  \caption{\label{fig::3loop}Left: Real (red) and imaginary (green) 
    parts of the three-loop light-fermion contributions to the factor 
    $F_{1}$ as a function of $\sqrt{s}$.
    Right: Two-loop results to the Higgs-gluon-gluon form factor 
    (see Ref.~\cite{Davies:2024znp} for a precise definition) with an
    off-shell gluon with virtuality $q_s^2$.}
\end{figure}

A straightforward extension of the two-loop calculation of
Ref.~\cite{Davies:2023vmj} (cf. Section~\ref{sec::nlo}) are the light-fermion
contributions, see Fig.~\ref{fig::3loop_class}(a).  In
Ref.~\cite{Davies:2023obx} we have performed a proof-of-principle calculation
and have computed the form factors for $t=0$ and $m_H=0$.  At one- and
two-loop order this approximation works a the level of 20\% or better for
$p_T=100$~GeV.  At three-loop order this would be sufficient to reduce the
uncertainties from the top quark mass scheme.
In Fig.~\ref{fig::3loop} we show the light-fermion contribution of $F_1$ for
$t=0$ and $m_H=0$.

The findings of Ref.~\cite{Davies:2023obx} show that the NNLO results
can be obtained in the forward scattering approximation
with the main bottleneck being the reduction to master integrals.
In the meantime we have performed the reduction of all
integrals for $t=0$ and $m_H=0$ contributing to the class
in Fig.~\ref{fig::3loop_class}(c). For the most complicated family 
the reduction took about 40~days requiring more than 2~terabytes
of RAM. In a first step we arrive at more than 30.000 master
integrals. For their minimization, which also involves integration-by-parts
reductions, we use a refined version of the approach
described in Ref.~\cite{Davies:2018qvx}. Here the command {\tt FindRules}
implemented in {\tt FIRE}~\cite{Smirnov:2023yhb} is helpful. 

The approach described above can also be applied to
parts of the calculation of the diagrams in Fig.~\ref{fig::3loop_class}(d).
This class of diagrams has the additional complication that
there are asymptotic contributions beyond the Taylor expansion which need
additional investigations.

Results for finite Higgs boson mass which are valid in the whole phase space
have been obtained for the class of diagrams in
Fig.~\ref{fig::3loop_class}(b)~\cite{Davies:2024znp}. The calculation
factorizes into one- and two-loop form factors with an off-shell external
gluon with virtuality $q_s^2$. We have computed this contribution for a finite
final-state Higgs boson mass $m_H$ by applying expansions both for small
$q_s^2$ and small $m_H$.  The right panel in Fig.~\ref{fig::3loop}
demonstrates that the expansions (represented by different colours) have
sufficient overlap such that the whole physical region can be covered. The
contributions in Fig.~\ref{fig::3loop_class}(d) are neither separately finite
nor gauge parameter independent. In both cases the contributions from
Fig.~\ref{fig::3loop_class}(d) are needed.

%
%
%
%


\section*{Acknowledgements}  

This research was supported by the Deutsche Forschungsgemeinschaft (DFG,
German Research Foundation) under grant 396021762 --- TRR 257 ``Particle
Physics Phenomenology after the Higgs Discovery''.  I would like to thank the
organizers of ICHEP2024 for the posibility to present our results at the
conference. Furhtermore, I would like to thank Joshua Davies, Go Mishima, Kay
Sch\"onwald and Marco Vitti for the fruitful collaboration on the topics
discussed in this contribution.


\bibliographystyle{jhep} 
\bibliography{inspire.bib}

%

\end{document}